\documentclass[prl,twocolumn,superscriptaddress]{revtex4}
\usepackage{amsmath}
\usepackage{amsfonts}
\usepackage{mathrsfs}
\usepackage{graphicx}
\usepackage{times}
\usepackage{color}
\usepackage{multirow}
\usepackage{dcolumn}% Align table columns on decimal point
\usepackage{bm}% bold math
\usepackage{amsmath}
\usepackage{amssymb}
\usepackage{amsfonts}
\usepackage{multirow}
\begin{document}

\title{Experimental investigation of the uncertainty relation in pre- and postselected systems}

\author{Yue Zhang}
\thanks{These authors contribute equally to this work.}
\affiliation{Ministry of Education Key Laboratory for Nonequilibrium Synthesis and Modulation of Condensed Matter,Shaanxi Province Key Laboratory of Quantum Information and Quantum Optoelectronic Devices, School of Physics, Xi'an Jiaotong University, Xi'an 710049, China}

\author{Xinyang Che}
\thanks{These authors contribute equally to this work.}
\affiliation{Ministry of Education Key Laboratory for Nonequilibrium Synthesis and Modulation of Condensed Matter,Shaanxi Province Key Laboratory of Quantum Information and Quantum Optoelectronic Devices, School of Physics, Xi'an Jiaotong University, Xi'an 710049, China}

\author{Yuanbang Wei}
\affiliation{Ministry of Education Key Laboratory for Nonequilibrium Synthesis and Modulation of Condensed Matter,Shaanxi Province Key Laboratory of Quantum Information and Quantum Optoelectronic Devices, School of Physics, Xi'an Jiaotong University, Xi'an 710049, China}

\author{Rui Tian}
\affiliation{Ministry of Education Key Laboratory for Nonequilibrium Synthesis and Modulation of Condensed Matter,Shaanxi Province Key Laboratory of Quantum Information and Quantum Optoelectronic Devices, School of Physics, Xi'an Jiaotong University, Xi'an 710049, China}

\author{Yi-an Li}
\affiliation{Ministry of Education Key Laboratory for Nonequilibrium Synthesis and Modulation of Condensed Matter,Shaanxi Province Key Laboratory of Quantum Information and Quantum Optoelectronic Devices, School of Physics, Xi'an Jiaotong University, Xi'an 710049, China}

\author{Miao Zhang}
\affiliation{Ministry of Education Key Laboratory for Nonequilibrium Synthesis and Modulation of Condensed Matter,Shaanxi Province Key Laboratory of Quantum Information and Quantum Optoelectronic Devices, School of Physics, Xi'an Jiaotong University, Xi'an 710049, China}

\author{Shuai Li}
\email{lishuai0999@stu.xjtu.edu.cn}
\affiliation{Ministry of Education Key Laboratory for Nonequilibrium Synthesis and Modulation of Condensed Matter,Shaanxi Province Key Laboratory of Quantum Information and Quantum Optoelectronic Devices, School of Physics, Xi'an Jiaotong University, Xi'an 710049, China}

\author{Bo Liu}
\email{liubophy@gmail.com}
\affiliation{Ministry of Education Key Laboratory for Nonequilibrium Synthesis and Modulation of Condensed Matter,Shaanxi Province Key Laboratory of Quantum Information and Quantum Optoelectronic Devices, School of Physics, Xi'an Jiaotong University, Xi'an 710049, China}

\begin{abstract}
Uncertainty principle is one of the fundamental principles of quantum mechanics. Exploring such uncertainty relations in pre- and postselected (PPS) systems, where weak measurements on post-selected states have been used as a powerful tool for exploring the foundation of quantum mechanics, has so far eluded experimental effort. In this work, we experimentally investigate the Robertson-Heisenberg-type uncertainty relation for two incompatible observables in a PPS system. Through conducting a von Neumann-type measurement, uncertainty relations between two non-commuting Pauli observables are experimentally unveiled. Such uncertainty relations in the PPS system impose limitations on the joint sharp preparation of pre- and postselected states for two incompatible observables. Our experiments would be useful for the development of amplification techniques for precision measurements.
\end{abstract}

\maketitle

\section{\label{sec:level1}Introdution}
Uncertainty relations~\cite{ref1, ref2, ref3, ref4} play fundamental roles and have dominated the development of physics that ranges from foundations to applications, such as quantum metrology~\cite{ref15}, entanglement detection~\cite{ref16, ref17} and quantum cryptography ~\cite{ref18}. The study of uncertainty relations continues to be at the center stage even after many years of the development of quantum theory~\cite{ref5, ref6, ref7, ref8, ref9, ref10, ref11, ref12, ref13, ref14}. On one hand, there is the Robertson uncertainty relation~\cite{ref28} that is supposed to capture the preparation uncertainty about the quantum ensemble. On the other hand,
it can be discussed based on the measurement-disturbance principle~\cite{ref42, ref43, ref44, ref45, ref32, ref33, ref34, ref35, ref36, ref37, ref38, ref39, ref40, ref41}. The later principle tries to formalise the original thought experiment of Heisenberg quantitatively. In the standard quantum system, one of the most well-known uncertainty relations is the {Robertson-Heisenberg uncertainty relation} (RHUR), which can be expressed as
\begin{equation}
\langle \Delta A \rangle_{\psi}^2 \langle \Delta B \rangle_{\psi}^2 \geqslant \left[ \frac{1}{2i} \langle \psi | [A,B] | \psi \rangle \right]^2,
\label{eq:1}
\end{equation}
where expected values and corresponding variances over an arbitrary state $|\psi\rangle$ can be defined as \( \langle A(B) \rangle_{\psi} = \langle \psi | A(B) | \psi \rangle \) and \( \langle \Delta A(B) \rangle_{\psi} = \sqrt{\langle \psi | A^2(B^2) | \psi \rangle - \langle \psi | A(B) | \psi \rangle^2} \), respectively, with $A(B)$ being a certain observable. The commutator is defined as \([A,B] = AB - BA\).

Based on RHUR, a stronger extension of that, i.e., {Robertson-Shr\"odinger uncertainty relation}, is formulated as
\begin{equation}
\begin{split}
\langle \Delta A \rangle_{\psi}^2 \langle \Delta B \rangle_{\psi}^2 \geqslant \left[ \frac{1}{2i} \langle \psi | [A,B] | \psi \rangle \right]^2 \\
+ \left[ \frac{1}{2} \langle \psi | \{A,B\} | \psi \rangle - \langle A \rangle_{\psi} \langle B \rangle_{\psi} \right]^2,
\end{split}
\label{eq:2}
\end{equation}
where the anticommutator is defined as \(\{A,B\} = AB + BA\). However, the above two uncertainty relations fail to capture the incompatibility between non-commuting observables, since they could become trivial even when $A$ and $B$ are incompatible. To correct this problem, instead of using the product of variances, Maccone and Pati~\cite{ref19} proposed two stronger uncertainty relations based on the sum of variances. The first one can be expressed as
\begin{equation}
\langle \Delta A \rangle_\psi^2 + \langle \Delta B \rangle_\psi^2 \geqslant \pm i \langle \psi | [A, B] | \psi \rangle + |\langle \psi | (A \pm iB) | \psi^\perp \rangle|^2,
\label{eq:3}
\end{equation}
where the sign $\pm$ should be chosen so that $\pm i\langle \psi | [A, B] | \psi \rangle$ is positive. $|\psi\rangle$ is an arbitrary state and $|\psi^\perp\rangle$ is orthogonal to $|\psi\rangle$. The second uncertainty relation is
\begin{equation}
\langle \Delta A \rangle_\psi^2 + \langle \Delta B \rangle_\psi^2 \geqslant \frac{1}{2} \left| \langle \psi_{A+B}^\perp | (A + B) |\psi\rangle \right|^2,
\label{eq:4}
\end{equation}
where \( |\psi_{A+B}^\perp \rangle \propto (A + B - \langle A + B \rangle_\psi )|\psi\rangle \).

In parallel, the ideas of pre- and postselected systems were also developed from a conceptual point of view, leading to new studies, such as on exploring whether uncertainty relations will impose fundamental limitations on the preparation of the PPS ensembles while measuring two noncommuting observables during the weak measurement or not. Such concepts, like the weak value and PPS, were firstly introduced by Aharonov, Bergmann, and Lebowitz (ABL)~\cite{ref26} when investigating properties of a quantum system in PPS ensembles~\cite{ref27}. As distinct from the usual measurement, where the eigenvalue of the observable with certain probability and corresponding experimental error bars can be obtained, in the weak measurement of an observable, we do not get an eigenvalue, but a complex number, i.e., the weak value. Therefore, to formulate the uncertainty relation with the weak measurement of two non-commuting observables, we have to define the variance of an operator such that its average over the pre-selected (or post-selected) state gives us the weak value. Due to the remarkable properties of the weak value~\cite{ref28, ref29, ref30, ref31, ref32, ref33, ref34, ref35, ref36, ref37, ref38, ref39, ref40, ref41}, such as the weak value amplification, it has attracted a tremendous research interests~\cite{ref42, ref43, ref44, ref45, ref46, ref47, ref48}. Weak measurements in the PPS system have been used as a powerful tool in a wide range of contexts, including superluminal light propagations~\cite{ref31, ref46}, cavity QED~\cite{ref49} and precision measurements ~\cite{ref48, ref38, ref50, ref51}. However, in PPS systems, to explore uncertainty relations has so far eluded experimental effort.

In this work, we experimentally investigate the Robertson-Heisenberg-type uncertainty relation for two incompatible observables in PPS systems. A von Neumann-type measurement has been preformed in single photons in our experiments, which is used for constructing the uncertainty relation between two non-commuting Pauli observables experimentally. We find that our experimental results match well with the theoretical predictions.

\section{EXPERIMENTAL INVESTIGATION of the uncertainty relation in PPS systems}
We investigate the uncertainty relation of two incompatible observables by choosing two Pauli operators $\sigma_x$ and $\sigma_y$ for spin-$\frac{1}{2}$ particle as two observables: \begin{equation}
\sigma_x =
\begin{pmatrix}
0 & 1 \\
1 & 0 \\
\end{pmatrix}, \quad
\sigma_y =
\begin{pmatrix}
0 & -i \\
i & 0 \\
\end{pmatrix}.
\label{eq:10}
\end{equation}

\begin{figure}[htbp]
    \centering
    \includegraphics[width=\linewidth]{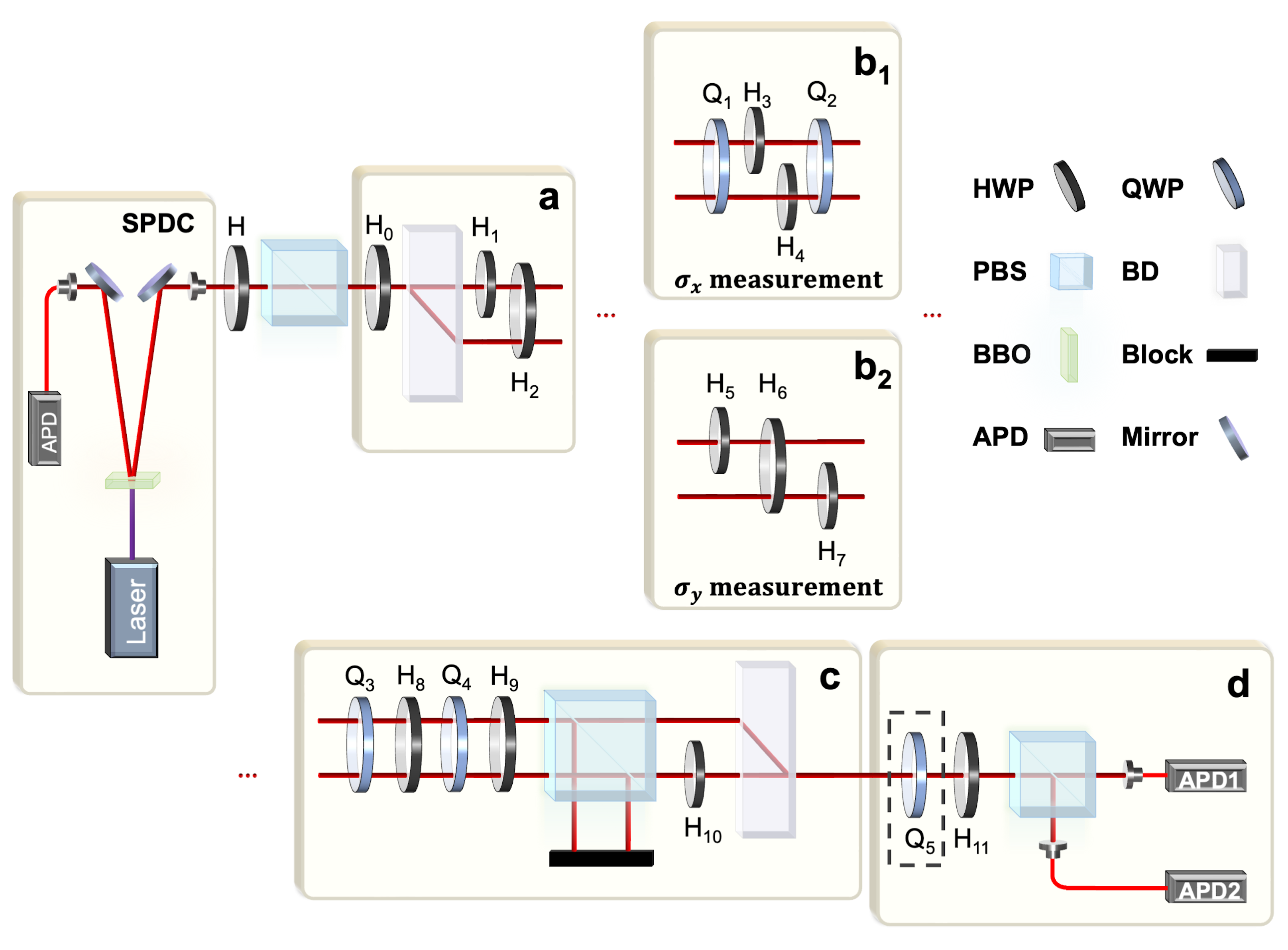}
    \caption{Experimental setup for investigating the uncertainty relation between two non-commuting Pauli observables in PPS systems. Photon pairs are produced via spontaneous parametric down conversion (SPDC). The signal photons are initially prepared in module a and then sent into weak measurement modules $b_{1}$ and $b_{2}$. In module $b_{1}$ and $b_{2}$, different wave plates are used in the weak measurement of $\hat{\sigma}_x$ and $\hat{\sigma}_y$, respectively. In module c, the system is post-selected to \( |\phi \rangle = \cos(\omega/2) |H\rangle + e^{i\eta}\sin(\omega/2) |V\rangle \) with \(\omega = \pi/3\) and \(\eta = \pi/5\). After performing projective measurements, coincidence counting of photons is performed by APDs in module d.}
    \label{fig:main}
\end{figure}
The resulting bound on the product of their variances in the PPS system~\cite{ref20} can be expressed as
\begin{equation}
\left(\langle \Delta \sigma_x \rangle_{\psi}^2 + \epsilon_{\sigma_x}\right) \left(\langle \Delta \sigma_y \rangle_{\psi}^2 + \epsilon_{\sigma_y}\right) \geqslant \left[\langle \sigma_z \rangle_{\psi} - \text{Im}(W_{\sigma_x \sigma_y}) \right]^2,
\label{eq:15}
\end{equation}
where \( \epsilon_{A(B)} = \langle A(B) \rangle_{\psi}^2 - |\langle A(B) \rangle_{\psi}^{\phi}|^2 |\langle \phi|\psi \rangle|^2 \) and \(W_{AB} = (\langle A \rangle_{\psi}^{\phi})^* \langle B \rangle_{\psi}^{\phi} |\langle \phi | \psi \rangle|^2\). Here, the weak value of the observable $A(B)$ is defined as \(\langle A(B) \rangle_{\psi}^{\phi} = \langle \phi | A(B) | \psi \rangle / \langle \phi | \psi \rangle\) with \( A(B) = \sigma_x \) or \( \sigma_y \). \( |\psi\rangle \) and \( |\phi\rangle \) are pre- and postselected states, respectively.

Adding ``Schr\"odinger's'' term, Eq.~(\ref{eq:15}) becomes
\begin{equation}
\begin{aligned}
    \left(\langle \Delta \sigma_x \rangle_{\psi}^2 + \epsilon_{\sigma_x}\right)
    \left(\langle \Delta \sigma_y \rangle_{\psi}^2 + \epsilon_{\sigma_y}\right)
    &\geqslant
    \left[\langle \sigma_z \rangle_{\psi} - \text{Im}(W_{\sigma_x \sigma_y}) \right]^2 \\
    &\quad + \left[\text{Re}(W_{\sigma_x \sigma_y}) \right]^2,
\end{aligned}
\label{eq:16}
\end{equation}

Besides the above two uncertainty relations, the uncertainty relation based on the sum of variances can also be obtained
\begin{equation}
    \left(\langle \Delta \sigma_x \rangle_{\psi}^2 + \epsilon_{\sigma_x}\right) + \left(\langle \Delta \sigma_y \rangle_{\psi}^2 + \epsilon_{\sigma_y}\right)
    \geqslant
    \mp (2 \langle \sigma_z \rangle_{\psi}+2\text{Im}(W_{\sigma_x \sigma_y})),
\label{eq:17}
\end{equation}
Here, the $\mp$ sign is taken suitably such that the right side of Eq. (8) is always positive.

To experimentally investigate the above uncertainty relations in Eq.~(\ref{eq:15}--\ref{eq:17}), we conduct a von Neumann-type measurement~\cite{ref22, ref23} to obtain the weak value of observables $\sigma_x$ and $\sigma_y$. Such a weak measurement can be realized using the interaction between the system and a pointer, where they can evolve together. Let us take the weak measurement of the system observable $\sigma_x$ as an example. The state evolution can be described as follows
\begin{align}
|\Psi_T\rangle &= \frac{1}{\sqrt{2}}(e^{-i\hat{\sigma}_x \otimes \gamma \hat{\tau}_z} |\psi_i\rangle |0\rangle_p-e^{i\hat{\sigma}_x \otimes \gamma \hat{\tau}_z} |\psi_i\rangle |1\rangle_p) \nonumber \\
&= \frac{1}{\sqrt{2}}[(\cos \gamma - i \hat{\sigma}_x \otimes \hat{\tau}_z \sin \gamma) |\psi_i\rangle |0\rangle_p -(\cos \gamma \nonumber \\
&+ i \hat{\sigma}_x \otimes \hat{\tau}_z \sin \gamma) |\psi_i\rangle |1\rangle_p],
\end{align}
where the interaction Hamiltonian is captured by \(\gamma \hat{A} \otimes \hat{\tau}_z \) with \( \gamma \) representing the coupling constant and \( \hat{A} = \hat{\sigma}_{x} \) standing for the observable of the system. Here, the pointer is described by \(\hat{\tau}_z = \begin{pmatrix} 1 & 0 \\ 0 & -1 \end{pmatrix}\) and its qubit space is spanned by the orthogonal states \( |0\rangle_p\) and \( |1\rangle_p\).
\(|\psi_i\rangle\) is the initial state of the system. Since the system observables are the Pauli-type~\cite{ref24}, we use the following relation \(e^{\pm i\hat{\sigma}_x \otimes \gamma \hat{\tau}_z} = \cos \gamma \pm i \hat{\sigma}_x \otimes \hat{\tau}_z \sin \gamma\) to simplify the Eq. (9). After post-selecting the system into the state \( |\phi\rangle \), the pointer state becomes
\begin{align}
|\tilde{\varphi}_p\rangle &= \langle \phi | \Psi_T \rangle \nonumber \\
&=\frac{1}{\sqrt{2}} \langle \phi | \psi_i \rangle [( \cos \gamma - i \sin \gamma \langle \hat{\sigma}_x\rangle_{\psi_i}^{\phi})|0\rangle_p \nonumber \\
&+( \cos \gamma + i \sin \gamma \langle \hat{\sigma}_x\rangle_{\psi_i}^{\phi})|1\rangle_p].
\end{align}
The real and imaginary part of \( \langle \hat{\sigma}_x \rangle_{\psi_i}^{\phi} \) can be obtained from the expectation values of pointer observables \( \hat{\tilde{{\sigma}}}_y \) and \( \hat{\tilde{{\sigma}}}_z \) through the following relations
\begin{equation}
\langle \hat{\tilde{{\sigma}}}_y \rangle_p = \frac{\text{Re}(\langle \hat{\sigma}_x \rangle_{\psi_i}^{\phi}) \sin 2\gamma}{\cos^2 \gamma + |\langle \hat{\sigma}_x \rangle_{\psi_i}^{\phi}|^2 \sin^2 \gamma},
\end{equation}
\begin{equation}
\langle \hat{\tilde{{\sigma}}}_z \rangle_p = \frac{\text{Im}(\langle \hat{\sigma}_x \rangle_{\psi_i}^{\phi} )\sin 2\gamma}{\cos^2 \gamma + |\langle \hat{\sigma}_x \rangle_{\psi_i}^{\phi}|^2 \sin^2 \gamma},
\end{equation}
where \( \hat{\tilde{{\sigma}}}_z = \begin{pmatrix}
1 & 0 \\
0 & -1 \\
\end{pmatrix}\), \( \hat{\tilde{{\sigma}}}_y = \begin{pmatrix}
0 & -i \\
i & 0 \\
\end{pmatrix}\) and \( \langle \hat{\tilde{{\sigma}}}_{y (z)} \rangle_p = \langle \tilde{\varphi}_{p} | \hat{\tilde{{\sigma}}}_{y (z)} | \tilde{\varphi}_{p} \rangle / \langle \tilde{\varphi}_{p} | \tilde{\varphi}_{p} \rangle \).

Our experimental setup is shown in Fig.~\ref{fig:main}. In experiments, the single-photon source is generated through pumping a $\beta$-barium-borate (BBO) nonlinear crystal via a CW diode laser. Polarization-degenerate photon pairs at 810nm are produced through the type-I spontaneous parametric down-conversion (SPDC) process. Using one photon as the trigger, the signal photon is heralded in the measurement setup. We monitor such trigger-signal photon pairs by a coincidence count at two APDs. The total coincidence counts are about $10^4$ over a collection time of 1s.

\begin{table}[h!]
\caption{The angles of wave plates for the weak measurement in our experiments.}
\centering
\renewcommand{\arraystretch}{1.5} % Increases the vertical spacing between rows
\begin{ruledtabular}
\begin{tabular}{c c c c c c c c}
\textit{H$_0$} & \textit{H$_1$} & \textit{H$_2$} & \textit{H$_3$} & \textit{H$_4$} & \textit{H$_5$} &
\textit{H$_6$} & \textit{H$_7$} \\
\hline
22.50$^\circ$ & 45.00$^\circ$ & $\theta/4$ & -22.50$^\circ$ & 22.50$^\circ$ & 0.00$^\circ$ & 22.50$^\circ$ & 0.00$^\circ$\\
\hline
\textit{H$_8$} & \textit{H$_9$} & \textit{H$_{10}$} & \textit{Q$_1$} & \textit{Q$_2$} & \textit{Q$_3$} & \textit{Q$_4$} \\
\hline
-9.00$^\circ$ & -15.00$^\circ$ & 45.00$^\circ$ & 0.00$^\circ$ & 0.00$^\circ$ & 45.00$^\circ$ & 45.00$^\circ$ \\
\end{tabular}
\end{ruledtabular}
\label{tab:waveplates}
\end{table}

\begin{table}[h!]
\caption{The angles of \textit{Q$_5$} and \textit{H$_{11}$} for the measurement of the observable $\hat{\sigma}_x$. Here, "--" indicates that the corresponding wave plate or APD was removed. "\checkmark" denotes the active measurement device. For the $\hat{\sigma}_y$ measurement, \( n_1 \) and \( n_4 \) denote the photon coincidence at APD2. Other experimental sets are the same as in the measurement of $\hat{\sigma}_x$.}
\centering
\renewcommand{\arraystretch}{1.5}
\begin{ruledtabular}
\begin{tabular}{c c c c c}
measurement & \textit{Q$_5$} & \textit{H$_{11}$} & APD1 & APD2 \\
\hline
\textit{n$_1$} & 0.00$^\circ$ &  22.50$^\circ$ & \checkmark & -- \\

\textit{n$_2$} & -- & 22.50$^\circ$ & -- & \checkmark \\

\textit{n$_3$} & -- & 22.50$^\circ$ & \checkmark & -- \\

\textit{n$_4$} & -- & 0.00$^\circ$ & \checkmark & -- \\

\textit{n$_5$} & -- & 0.00$^\circ$ & \checkmark & -- \\
\end{tabular}
\end{ruledtabular}
\label{tab:measurements}
\end{table}

The state initialization is conducted in the unit as shown in Fig.~1a. After the signal photons pass through a polarizing beam splitter (PBS) and a half-wave plate (HWP), where the angle of $H_0$ is fixed at \( \pi/8 \), the photons are separated into two parallel spatial modes by a beam displacer (BD), where vertically polarized photons \( |V\rangle \) are transmitted directly, while horizontally polarized photons \( |H\rangle \) are displaced into the adjacent spatial mode. Two HWPs are then applied, where the angles of $H_1$ and $H_2$ are set at \( \pi/4 \) and $\theta/4$, respectively. In our experiments, we choose \(\theta = j\pi/9\) (\textit{j} = -9, \dots, 9), a total of 19 states for studying the uncertainty relation. In our experimental realization of the weak measurement, the pointer is encoded into the path degrees of freedom (DOF). And the polarization DOF is used for the system. The unitary evolution, for instance, associated with the system observable $\hat{\sigma}_x$, as shown in Fig.~1$b_{1}$, is achieved through using two quarter-wave plates (QWPs), where the angles of $Q_1$ and $Q_2$ are set at \textit{0}, and two half-wave plates (HWPs) with angles of \(H_3\) and \(H_4\) set at \(-\gamma/2\) and \(\gamma/2\) degrees, respectively. In our experiments, \(\gamma\) is set as \( \pi/4 \).

After post-selecting the system, realized in the experimental unit as shown in Fig.~1c, we measure the pointer observables \( \hat{\tilde{{\sigma}}}_z \) and \( \hat{\tilde{{\sigma}}}_y \) to extract the weak values. For instance, when measuring  the weak value of $\hat{\sigma}_x$, we first set the angles of $Q_5$ and $H_{11}$ at \textit{0} and $\pi/8$, respectively. The photon coincidence measured by APD1 is recorded as $n_1$. Then, we remove $Q_5$, and the photon counts at APD1 and APD2 are recorded as $n_3$ and $n_2$, respectively. We then set the angle of $H_{11}$ at \textit{0} and the measured photon coincidence at APD1 is denoted as $n_4$. Finally, we remove set $b_{1}$ and perform the projective measurement with $H_{11}$ set at \textit{0}, obtaining the photon coincidence $n_5$ at APD1. The real and imaginary parts of the weak value of the observable $\sigma_x$ can be obtained through the following relations
$\frac{n_2 + n_3 - 2n_1}{4 \sin \gamma \cos \gamma n_5}$ and $\frac{2n_4 - n_2 - n_3}{4 \sin \gamma \cos \gamma n_5}$, respectively.

\begin{figure}[htbp]
\centering
\includegraphics[width=\columnwidth]{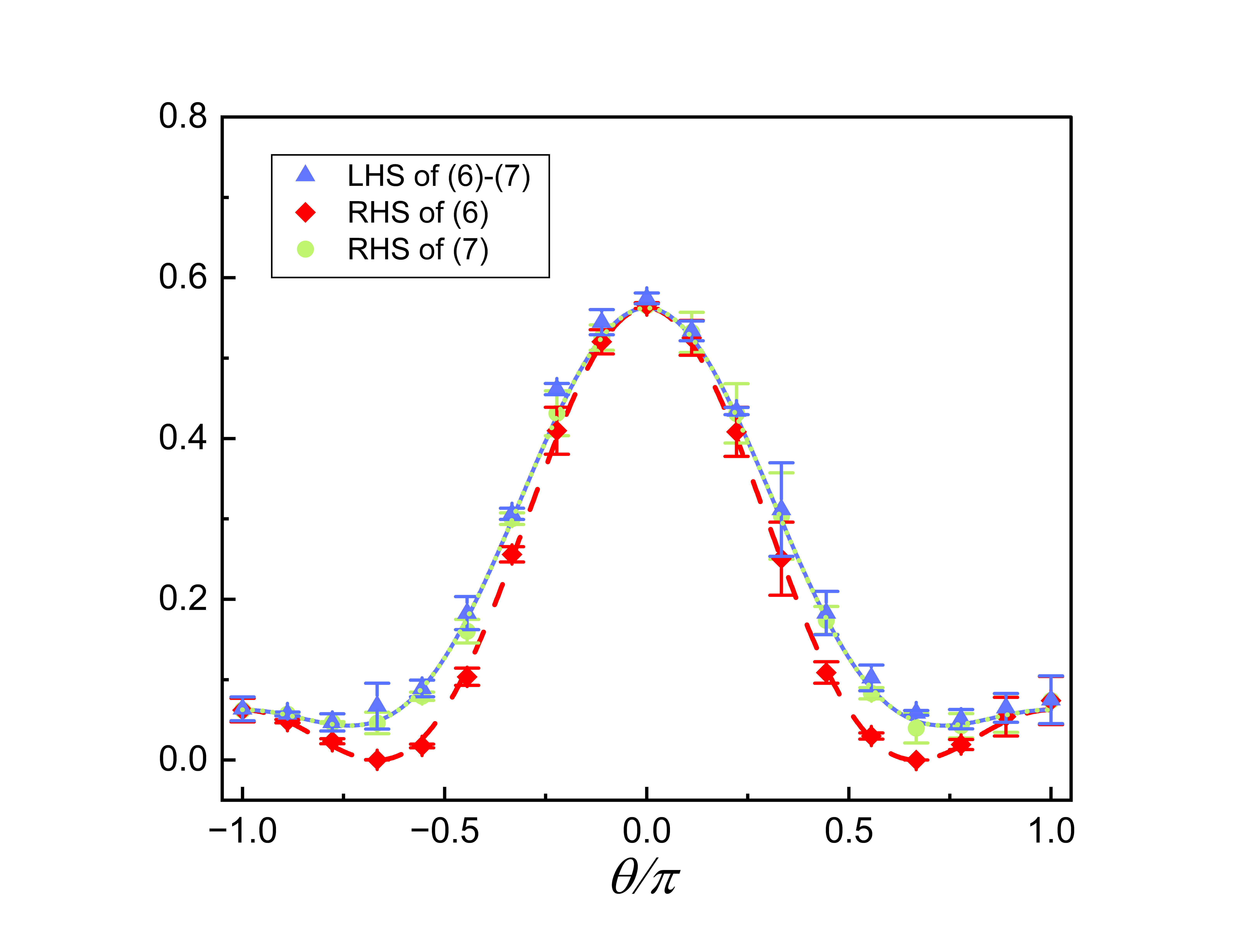}
\caption{Experimental investigation of the uncertainty relation in Eq.~(\ref{eq:15}) and Eq.~(\ref{eq:16}). Blue triangles and solid line represent the experimental results and theoretical predictions of the left-hand side (LHS) of Eq.~(\ref{eq:15}) and Eq.~(\ref{eq:16}), respectively. Red diamonds and dashed line show the experimental results and theoretical predictions of the right-hand side (RHS) of Eq.~(\ref{eq:15}). Green dots and dotted line represent the experimental results and theoretical predictions of the RHS of Eq.~(\ref{eq:16}). Here, we choose 19 different initial states through changing $\theta$.}
\label{fig:product}
\end{figure}

\begin{figure}[htbp]
\centering
\includegraphics[width=\columnwidth]{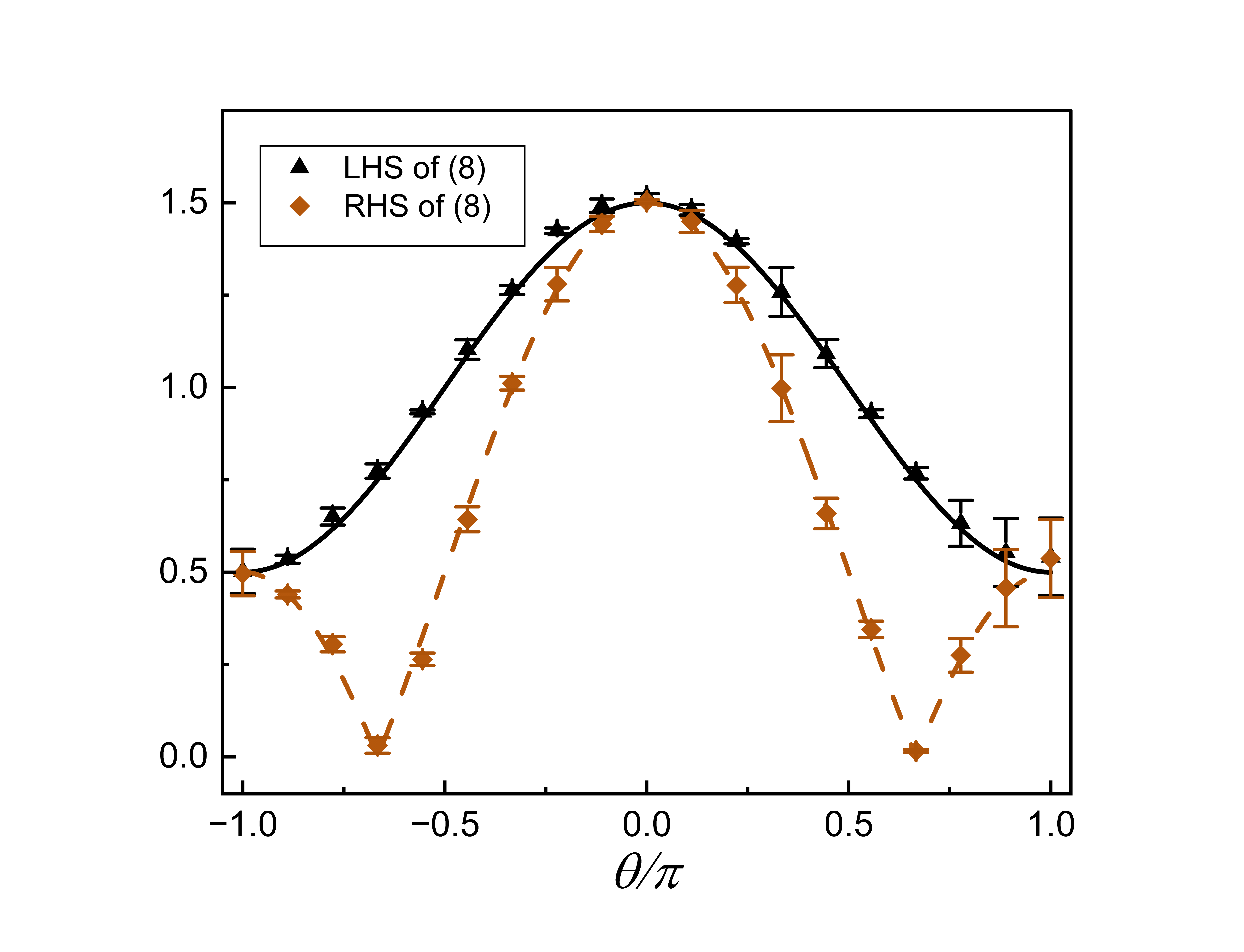}
\caption{Experimental investigation of the uncertainty relation in Eq.~(\ref{eq:17}) based on the sum of variances of two incompatible Pauli observables in PPS systems.
Black triangles and solid line represent the experimental results and theoretical predictions of the LHS of Eq.~(\ref{eq:17}), respectively. Brown diamonds and dashed line show the experimental results and theoretical predictions of the RHS of Eq.~(\ref{eq:17}). Here we also choose 19 different initial states through changing $\theta$.}
\label{fig:sum}
\end{figure}

As shown in Fig.~\ref{fig:product}, we show the experimental investigations of the uncertainty relations (\ref{eq:15}) and (\ref{eq:16}), based on the product of variances of two non-commuting Pauli observables in PPS systems. It is shown that experimental results agree with the theoretical predictions well. Fig.~\ref{fig:product} shows that the relation (\ref{eq:16}) is stronger than that in Eq.~(\ref{eq:15}) since it is nontrivial for all the values of $\theta$. We also show the uncertainty relation (\ref{eq:17}) based on the sum of variances of two incompatible Pauli observables in PPS systems in Fig.~\ref{fig:sum}. It is shown that our experimental data matches well with the theoretical prediction. Therefore, through conducting weak measurements in PPS systems, both product and sum uncertainty equalities for two non-commuting Pauli observables are investigated experimentally.

\section{CONCLUSION} In summary, we have experimentally investigated the Robertson-Heisenberg-type uncertainty relation for two incompatible observables in PPS systems. Relying on constructing a stable interferometric network with simple linear optical elements, a von Neumann-type measurement has been preformed in single photons in our experiments. Such weak measurements are utilized for experimentally studying the uncertainty relations between two non-commuting Pauli observables in PPS systems. It is shown that our experimental results match well with the theoretical prediction. Our study would shed light on the problem of the joint sharp preparation of pre- and postselected states for two incompatible observables. It thus can provide a correct understanding of the fundamental limitation of precise measurements in PPS systems.

\section{Acknowledgment}
This work is supported by the National Key R$\&$D Program of China (2021YFA1401700), NSFC (Grants No. 12074305, 12474267), the Fundamental Research Funds for the Central Universities (Grant No. xtr052023002), the Shaanxi Fundamental Science Research Project for Mathematics and Physics (Grant No. 23JSZ003) and the Xiaomi Young Scholar Program.

\bibliographystyle{apsrev}
\bibliography{apssamp}

\end{document}